\documentclass[ammsymb,pre,aps,twocolumn,superscriptaddress,showpacs]{revtex4}
\usepackage{graphicx}

\def \beq{\begin{equation}}
\def \eeq{\end{equation}}
\def \beqar{\begin{eqnarray}}
\def \eeqar{\end{eqnarray}}

\begin{document}

\title{Two classes of bipartite networks: nested biological and social
systems}
\author{Enrique Burgos}
\affiliation{Departamento de F{\'{\i}}sica, Comisi{\'o}n Nacional de Energ{\'\i }a At{\'o}%
mica,\\
Avenida del Libertador 8250, 1429 Buenos Aires, Argentina}
\affiliation{Consejo Nacional de Investigaciones Cient\'{\i}ficas y T\'{e}cnicas,\\
Avenida Rivadavia 1917, C1033AAJ, Buenos Aires, Argentina}
\author{Horacio Ceva}
\affiliation{Departamento de F{\'{\i}}sica, Comisi{\'o}n Nacional de Energ{\'\i }a At{\'o}%
mica,\\
Avenida del Libertador 8250, 1429 Buenos Aires, Argentina}
\author{Laura Hern\'{a}ndez}
\affiliation{Laboratoire de Physique Th\'eorique et Mod\'elisation; \\
UMR CNRS, Universit\'e de Cergy-Pontoise, \\
2 Avenue Adolphe Chauvin, 95302, Cergy-Pontoise Cedex France}
\author{R.P.J. Perazzo}
\affiliation{Departamento de Investigaci\'{o}n y Desarrollo, Instituto Tecnol\'{o}gico de
Buenos Aires \\
Avenida E. Madero 399, Buenos Aires, Argentina}
\author{Mariano Devoto}
\affiliation{C\'{a}tedra de Bot\'{a}nica, Facultad de Agronom\'{\i}a, Universidad de
Buenos Aires\\
Avenida San Mart\'{\i}n 4453, C1417DSE Buenos Aires, Argentina}
\author{Diego Medan}
\affiliation{C\'{a}tedra de Bot\'{a}nica, Facultad de Agronom\'{\i}a, Universidad de
Buenos Aires\\
Avenida San Mart\'{\i}n 4453, C1417DSE Buenos Aires, Argentina}
\affiliation{Consejo Nacional de Investigaciones Cient\'{\i}ficas y T\'{e}cnicas,\\
Avenida Rivadavia 1917, C1033AAJ, Buenos Aires, Argentina}
\date{\today}

\begin{abstract}
Bipartite graphs have received some attention in the study of social
networks and of biological mutualistic systems. A generalization of a
previous model is presented, that evolves the topology of the graph in order
to optimally account for a given Contact Preference Rule between the two
guilds of the network. As a result, social and biological graphs are
classified as belonging to two clearly different classes. Projected graphs,
linking the agents of only one guild, are obtained from the original
bipartite graph. The corresponding evolution of its statistical properties
is also studied. An example of a biological mutualistic network is analyzed
in detail, and it is found that the model provides a very good fitting of
all the main statistical features. The model also provides a proper
qualitative description of the same features observed in social webs,
suggesting the possible reasons underlying the difference in the
organization of these two kinds of bipartite networks.
\end{abstract}

\maketitle


\section{Introduction\label{seccion_Introduction}}

Bipartite networks have attracted considerable attention \cite{Newman}, \cite%
{physrep} because they can describe social and ecological systems. These
involve nodes of two kinds, and their edges only link nodes of different
guilds.

The examples issuing from biology concern ecological systems. These are
complex ensembles of living beings sharing a complicated pattern of mutual
dependence and interacting in many intricate ways. A number of these systems
provide valuable services to mankind and considerable attention is currently
being paid to their stability taking into account human disturbance. A
sustainable management of ecosystems can only be achieved if a proper
understanding is reached concerning how these systems are assembled. As far
as biological systems are concerned we will discuss the case of mutualistic
systems. They involve two groups of species, usually animals and plants,
that interact to fulfill essential biological functions such as feeding or
reproduction. This is the case of systems involving plants and animals that
feed from the fruits and disperse their seeds (\textit{seed dispersal
networks}). Another example is that of insects that feed from the nectar of
flowers while pollinating them in the process (\textit{pollination networks}%
).

Bipartite networks can also be found in social systems. Examples of this
type involve the actors and movies they participate in \cite{W&S} or the
boards of directors of large companies and their members \cite{Newman}.

An important feature of bipartite networks is the degree distributions of
the nodes of both guilds. In social systems the statistical properties of
the degree distributions of each guild are different. While the distribution
associated to one guild approximately follows a power law, the degrees of
the other distribute themselves as a bell shaped (Poisson-like) curve around
some average value. In biological systems the degree distributions of \emph{%
both} guilds decay slower than exponentially thus having fat tails. In spite
of the fact that observed mutualistic systems are rather small, these
distributions have been fitted by truncated power laws.

In addition it has been observed \cite{AyP}, \cite{muchas} that in
biological, mutualistic networks all the contacts (links) tend to be \emph{%
nested} and limited by a curve\cite{AyP}, \cite{nosotros1} defined as an 
\textit{isocline of perfect order} \cite{pie1}.

In a nested network the nodes of both types can be ordered by decreasing
degree in such a way that the set of species linked with each species in the
list is contained in the set associated to the preceding one. This
organization is such that the \textit{generalists} of both type of guilds
(i.e. those nodes interacting with a great number of nodes of the other
guild) tend to interact among them while there are no contacts among \textit{%
specialists} (i.e. nodes interacting with very few of the other guild). All
these features indicate that these networks are far from being a random
collection of interacting species, displaying instead a high degree of
internal organization.

In preceding papers we have introduced the Self-organizing Network Model
(SNM) \cite{nosotros1}, \cite{nosotros2} to describe nested biological webs.
Within that model, the topology of the network is the result of a
self-organization process in which its nodes progressively redefine their
links obeying to a purely local rule that does not depend upon any global
feature of the network. The network undergoes in this fashion an ordering
process.

This self-organization process by no means represents any growth or
assembling process of the ecological system. It represents instead the
search of a pattern of contacts between both guilds that tends to optimally
take into account some \emph{contact preference rule} (CPR) that is assumed
to prevail among the nodes of the network. The aim of this model is to trace
a possible causal relationship between the detailes of a local rule
governing the contacts between both guilds and the statistical features of
the network. In fact, the results obtained in \cite{nosotros1} under one
specific assumption for such CPR accurately account for many observed
statistical features of real mutualistic webs of widely different sizes.

In this article we present a generalization of the SNM to cast into a single
framework the organization of both, mutualistic and social webs. We thus aim
at establishing similarities and differences between social and biological
networks by linking their statistical properties with the CPR that governs
the interactions between both guilds.

The fact that nodes of one guild share contacts with nodes of the other
kind, allows us to define a pattern of interactions among similar nodes \cite%
{Newman}. Any bipartite network can also be regarded as describing two
separate systems involving only nodes of one kind. The corresponding \emph{%
projected graphs} are built by defining two nodes of the same guild as
neighbors - therefore linked by an edge - when both share a contact with at
least one node of the other guild. For instance, two plants are considered
as neighbors if they are visited by the same animal species or two directors
are neighbors if they belong to the same board.

Two nodes of one guild may share more than one contact with the nodes of the
other guild. As a consequence, in the projected graphs not all edges have
the same importance. Each edge carries a weight representing the number of
common neighbors of the other kind thus providing a measure of the intensity
of the corresponding interaction.This is the case for instance when more
than one animal species visit the same pair of plants or two actors
participate together in several films.

We also address the interesting question of how the gradual changes that are
involved in the SNM are reflected both in the topology and the weights of
the interactions in the projected graphs.

The aim of the previous analysis is to extract salient features of social
and biological networks and not to provide precise fittings to empirical
data. Nevertheless we compare in some detail theoretical predictions with
the values observed in a real mutualist web \cite{Robertson} which is one of
the largest mutualistic system reported in the literature, therefore
allowing for statistical considerations.

The theoretical discussions presented in this work shed light on several
questions. In the first place they allow to establish links between
nestedness and the shape of degree distributions of biological networks, in
the second place they provide hints about possible reasons for the different
shapes of the degree distributions of both guilds found in social networks.
These two elements place biological and social webs into well differentiated
classes of bipartite networks. Finally the theoretical predictions of the
SNM are extended to the projected graphs and are found to provide a faithfull 
description of the distributions observed in the mutualistic network of 
Ref. \cite{Robertson}.

\section{Theoretical background}

\subsection{The Self-organizing Network Model}

We describe here a generalization of the SNM introduced in Ref. \cite%
{nosotros1}. We refer the reader to that article for the details of the
original model as well as for the comparison of its results with the
empirical observations.

The interaction pattern of a bipartite network can be coded as an adjacency
matrix in which rows and columns are labeled respectively by the plant and
animal species involved in the network. Its elements $K_{p,a}\in \{0,1\}$
represent respectively the absence or the presence of an interaction between
the plant species $p$ and the animal species $a$. In what follows we drop
the term species specifying that when mentioning plants or animals we are
not referring to the behavior of separate individuals but to all the members
of a species.

The SNM is a computer model that starts from a random adjacency matrix in
which the number of plants, animals and contacts between them are
arbitrarily fixed provided that there are no species left without links with
the other guild. Starting from this initial configuration plants and animals
iteratively redefine their contacts by reallocating the 1's of the adjacency
matrix. This reallocation obeys to some assumed CPR. In the following we
consider a CPR that indicates that the agents of either guild prefer to set
contacts with a species of the opposite guild having a greater (or lesser)
number of contacts.

We implement such swapping with the following algorithm. In each iteration
first a row and next a column are chosen at random. Once a row (column) has
been chosen its contacts are reallocated with probability $P_{r}$ ($P_{c}$).
Reallocation consists in choosing at random a 1 and 0 belonging to the same
row (column), and swapping them acording to a previously selected CPR that
we discuss below in some detail. The row (column) is left unchanged with
probability $1-P_{r}$ ($1-P_{c}$). In case that, upon swapping, a row or a
column would be left with no links, the reallocation is not produced. This
rule prevents the elimination of a node of the system as a consequence of
being left without interactions.

The two probability parameters $P_r$ and $P_c$ must not be considered as
independent because the only relevant differences appear when their ratio $R
= P_r/P_c$ is changed i.e. when columns and rows are updated with different
frequencies. We will consider in particular two limiting situations, one in
which $R\simeq 1$ and another in which $R \ll 1$ or $R \gg 1$. These
respectively correspond to a situation in which rows and columns are updated
with the same frequency or to the case in which rows (columns) are updated
much more frequently than columns (rows).

The swapping process is continued until the CPR that has been imposed is
optimally satisfied and no further swappings can take place. The network
reaches then a perfectly ordered phase \cite{pie2}.

One possible CPR [strategy (I)] is that the degree of the new partner must
be higher than the one of the previous partner. Within this CPR species of
either kind tend to be as generalists as possible. An alternative
possibility [Strategy (II)] is just the opposite, namely that the new
partner has fewer contacts than the previous one. In this case species tend
to be as specialists as possible. Strategy (I) bears some similarity with
the rule of preferential attachment of Ref. \cite{B-A}. This is a stochastic
attachment rule by which new nodes are added to a growing network attaching
to the existing nodes with a probability that is proportional to their
degree. There are however several important differences between preferential
attachment and our strategy (I).

In the first place the approach in Ref.\cite{B-A} deals with a population of
entities, that are represented by the nodes of the graph, that grows
constantly. Our model deals instead with a closed system in which new nodes
are not added. It consequently involves a change of the topology of a
network with a \emph{constant number} of nodes and links. In the second
place preferential attachment is clearly a non local process because the
particular attachment of a new node is governed by the degree distribution
of all the nodes of the network. Opposed to this, the present model follows
a purely local rule. The reallocation of contacts with both CPR's involves
only the information of the current and of the target nodes of its
counterparts in the bipartite network and has no relation whatsoever to any
global feature of the network. It could be thought of as one species of
animals changing its current choice as a consequence of the better
conditions offered by an alternative species of plants that is more highly
or poorly visited.

As it is well known the fact that the decision rule is local is an important
feature if the problem of the reallocation of all the contacts of the
network is cast into the form of the optimization problem of fulfilling a
given CPR. Within this particular framework the SNM can be regarded as an
heuristic solution for it.

\subsection{The projected graphs}

The information contained in a bipartite network can be used to construct
two separate graphs, each composed of nodes belonging to a single guild.
This is done extracting two projected graphs fulfilling the rule that two
nodes of the same guild are neighbors - and therefore linked by an edge - if
they share a contact with at least one node of the other type in the
bipartite network \cite{Newman} .

Let $K$ be the adjacency matrix with elements $K_{p,a} \in \{0,1\}$ denoting
the contacts between the plant $p$ and the animal $a$. $K^T$ is the
transposed of $K$. The two matrices 
\begin{eqnarray}
W^{P}_{p,p^{\prime}}&=&KK^T = \sum_a
K_{p,a}K^T_{a,p^{\prime}}(1-\delta_{p,p^{\prime}})  \nonumber \\
W^{A}_{a,a^{\prime}}&=&K^TK=\sum_p
K^T_{a,p}K_{p,a^{\prime}}(1-\delta_{a,a^{\prime}})  \label{pesos}
\end{eqnarray}
encode the weighted adjacency matrix of the projected graphs \cite{newman2}
for plants ($W^{P}$) and animals ($W^{A}$). The diagonal elements: 
\begin{eqnarray}
D^P(p)&=& \sum_a (K_{p,a})^2 \\
D^A(a)&=&\sum_p (K^T_{a,p})^2
\end{eqnarray}
that are canceled from the sums in Eq.(\ref{pesos}) are the degrees of the
plant and animal nodes in the bipartite graph. The non vanishing off
diagonal elements of $W^{A,P}$ carry the information of the number of
different paths linking two nodes of the same kind involving not more than
one node of the other guild. These weights could be interpreted as the
intensity of the interaction between such pair of species. A suitable
generalization of the concept of degree for weighted graphs is just the
total number of paths connecting some given node with all nearest
neighboring nodes of the same kind, namely: 
\begin{equation}
S^{A(P)}(i)= \sum_{j} (1-\delta_{i,j})W^{A(P)}_{i,j}  \label{fuerza}
\end{equation}
This is defined as the \emph{strength} \cite{physrep} of the node. It
provides a measure of the relevance of the species $i$ in the plant- or
animal systems. The usual degree of the $i-$th animal or plant in the
projected graph is denoted by $D_{\pi}^{P(A)}(i)$ and is given by the number
of non zero elements in each row of the matrices $W^{A(P)}$.

Besides the above distributions, the projected graphs can also be
characterized by the distribution of its clustering. The clustering $C_i$ of
the i-th node of any graph is defined (Ref.\cite{W&S}) as 
\begin{equation}
C_i=\frac{e_i}{k_i(k_i-1)/2}  \label{clustering}
\end{equation}
where $e_i$ is the number of edges among the neighbors of the $i-$th node
and $k_i$ is its degree. The clustering coefficient is the fraction of first
neighbors of a node that are themselves, neighbors among them.

\subsection{Properties of the projected graphs}

It is convenient to derive some analytical results for the properties of the
projected graphs under the changes in the topology of the original bipartite
graph. To this end we assume a perfectly ordered bipartite graph described
by an adjacency matrix with $m$ rows and $n$ columns and a given probability
of contacts $\phi $ and discuss the degree distributions of the projected
graphs.

We first consider an adjacency matrix with the same number of 1's in all its
rows. This is the case for the perfect order produced by SNM using the CPR
of strategy (II). As discussed below it also approximately represents the
situation of a random adjacency matrix.

Let $k=n\phi$ be the number of 1's in each row. The probability that any two
rows share no contacts with the same species of the other guild - and are
therefore \textit{not} neighbors in the projected graph - is: 
\begin{equation}
q_{n,k}=\frac{ {n-k \atopwithdelims() k}}
{{n \atopwithdelims()k}}=\frac{(n-k)!^2}{n!(n-2k)!} 
\label{denspro}
\end{equation}
therefore the probability that a given row has $\ell$ neighbors and hence
has degree $\ell$ in the projected graph is 
\begin{equation}
P_{\ell|m}={m-1 \atopwithdelims ()k}(1-q_{n,k})^{\ell}q_{n,k}^{m-1-\ell} \
\ ; \ \ \ell=1,2\dots,m-1
\end{equation}
The number of rows with degree $\ell$ is $N_{\ell|m}=mP_{\ell|m}$. The
average degree of the row-species in the projected graph therefore is 
\begin{equation}
N_m=\sum_{\ell=0}^{m-1}\ell N_{\ell|m}=m(m-1)(1-q_{n,k})  \label{disgrapro}
\end{equation}
Since $k=n\phi$ this degree distribution is fully specified by the
dimensions of the matrix and the probability of contacts $\phi$. A
completely symmetric argument can be made for the column-species changing $n$
by $m$.

The degree distribution for the opposite case, i.e. when the Strategy (I) is
used and the system reaches an asymptotic order of perfect nestedness is
simpler to obtain. If there are no species with no contacts and the pattern
of interactions is nested, there exists at least one species of each guild
that is a full generalist, i.e. has contacts with \emph{all} species of the
other kind. Under this condition all the species of each guild have contacts
with the generalist of the other guild and it is therefore a neighbor of all
the other species of the same type. Such perfectly nested system gives
therefore rise to two projected graphs that are \textquotedblleft tiny
worlds": all species are neighbors of each other\cite{bascolittlworl}.

The above Eq. \ref{denspro} can also be used to derive a close estimate of
density of contacts $\phi _{\pi }^{P,A}$ of the two projected graphs for
plants and animals, provided that the bipartite adjacency matrix is random.
This is 
\begin{equation}
\phi _{\pi }^{P}(n,\phi )=1-q_{n,k=n\phi }=1-\frac{(n-n\phi )!^{2}}{%
n!(n-2n\phi )!}  \label{densipro}
\end{equation}%
The density for the projected graph for animals $\phi _{\pi }^{A}(m,\phi )$
is obtained from Eq.\ref{densipro} by changing $n$ by $m$. This probability
of contacts between nodes of the same guild is a rapidly growing function of 
$\phi $, the probability of contacts in the rectangular adjacency matrix.
Hence, in general, even very sparse adjacency rectangular matrices give rise
to densely connected projected networks.

\section{Results}

\subsection{Results for the bipartite graphs}

In the following, if not stated otherwise, we will discuss numerical
examples concerning Strategy (I). The reason for this is two-fold: on one
hand strategy (I) has a greater biological significance and it has been
successfully used in Ref.\cite{nosotros1} to account for the degree
distributions of several observed mutualistic systems of a wide range of
sizes. On the other hand the ordered patterns emerging from Strategy (II)
can well be approximated by a random adjacency matrix. The reason for this
is simple. The use of this strategy leads to a different situation in which
all nodes of the same kind tend to have the same number of links \cite{pie3}%
. If contacts are randomly assigned all species have \emph{on average} the
same degree. Thus, the iterative ordering of the SNM only tends to produce a
sharper delta-like function in the degree distributions, centered at the
corresponding average number of links.

\begin{figure}[tbp]
\includegraphics[width=12cm]{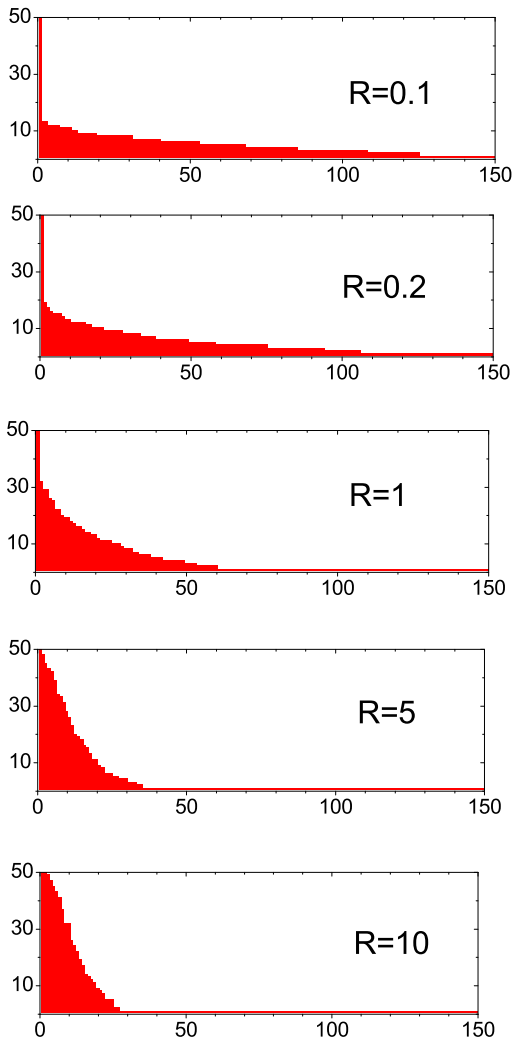}
\caption{{\protect\footnotesize Asymptotic adjacency matrices of a bipartite
network of $50 \times 150$. Each contact is shown as a black pixel. All
panels display the adjacency matrix obtained with the SNM using Strategy (I)
for both rows and columns and after $5 \times 10^6$ iterations.}}
\label{contactos}
\end{figure}

Whenever strategy (I) is used the model always leads to a perfectly nested
pattern, no matter the relative updating frequency of rows and columns, as
shown in Fig.\ref{contactos}. These perfectly ordered systems have been
obtained starting from a random adjacency matrix of $50 \times 150$ with a
probability of contacts between both guilds of 10\% and running the SNM
algorithm for a very large number of iterations until no further swappings
take place. The different panels of Fig. \ref{contactos} have different
values of $R$ and yet perfect nestedness is found in all cases.

The sole fact that interactions are arranged in a nested pattern does not
define the shape of the degree distributions. They do indeed differ
drastically with $R$. Several different shapes of this distribution can be
found that are nevertheless compatible with a nested pattern of interaction.
As we will soon see, the shape of the degree distribution actually provides
some information about the way in which the CPR is actually enforced among
the row- or column- species.

We discuss the numerical results obtained with the SNM by comparing them
with the real mutualistic system described in Ref.\cite{Robertson}. Real
systems such as this are not perfectly ordered. To obtain a theoretical
prediction from the SNM an initial configuration has to be chosen that
involves a random adjacency matrix with the same number of species and
interactions as the real system. The iterative ordering process starts from
this initial state and is stoped before a perfect order has been reached
using some appropriate stopping criterion that takes into consideration the
particular empirical situation under analysis. This is in fact the only
adjustable parameter of the model. The results shown here correspond to
100000 iterations of the SNM. In this case the stopping criterium is based
on a statistical estimate of the departure from the isocline of perfect
order. The algorithm is stopped when the value of this estimate is close to
the empirically observed one.

\begin{figure*}[tbp]
\includegraphics[width=18cm]{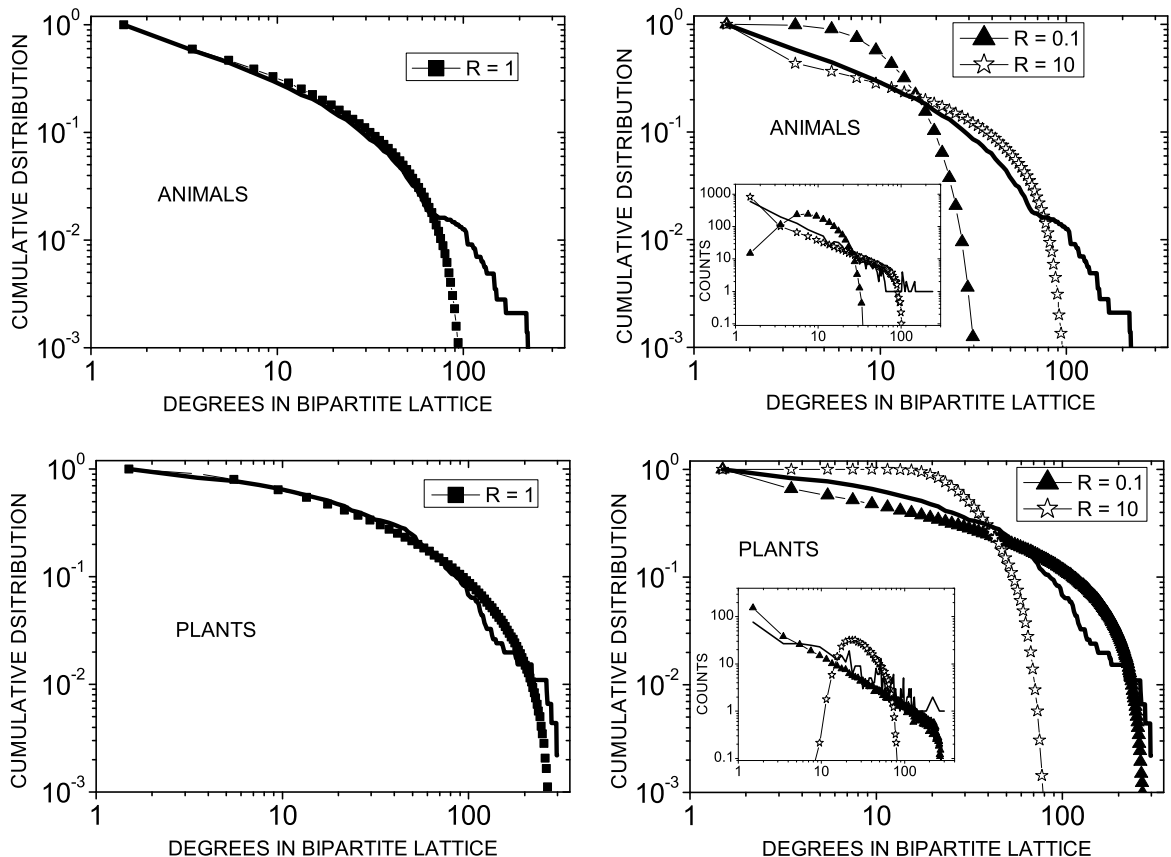}
\caption{{\protect\footnotesize Cumulative degree distributions of plants and 
animals of a bipartite network with the same dimensions ($456\times 1428$) as in Ref.%
\protect\cite{Robertson}, for different values of $R$. In the two panels on the right
the corresponding non cumulative distributions are shown in the insets. The empirical
distributions for plants and animals are always shown in heavy continuous line. 
Theoretical results are averaged over 100 realizations of the original random 
adjacency matrix.
All distributions have bin=2; this is the main reason for the noisy
appearance of the empirical data. Notice that in the plots of the right,
stars or triangles fail to account \textit{simultaneously} the empirical data. The non-cumulative
degree distribution shown in the insets illustrate the different organizations of the networks 
with $R=0.1$ and $R=10$ }}
\label{distribuciones}
\end{figure*}

In Fig.\ref{distribuciones} we show theoretical and empirically observed 
\cite{Robertson} cumulative degree distributions \cite{piex} for three 
values $R=1.0$, $R=0.1$ and $R=10.0$. The degree distributions for both 
guilds that are obtained with $R=1 $ and the above number of iterations, 
closely follow  truncated power laws, truncation being here a finite size 
effect.In the limit of perfect order and $R=1$ the
distributions for plants and animals can be shown to map into each other 
\cite{nosotros1} through the application of a simple scaling transformation.

The distributions for $R \ll 1$ or $R\gg 1$ for rows and columns shown in
the right panels of Fig. \ref{distribuciones} have quite different behaviors
and strongly depart from the observed distributions of the mutualistic
system. These curves show the seemingly paradoxical result that the degrees
of the guild that is updated \textit{less frequently} distribute according
to truncated a power law, while the \textit{more frequently} updated guild
has a distribution that is bell-shaped thus indicating that the distribution
of degrees have not been greatly changed by the self organization process
and have a distribution that resembles the original random pattern. This can
be understood because when, say, columns are frequently updated most
swappings take place within each column. The contacts that are changed are
therefore those of the row-agents while columns keep their degrees with
little change. Thus, the degree distribution of the rows changes while that
of the column-agents remains close to the original random matrix producing a
bell shaped curve. However this is not a transient-like behavior by which a
power law could be reached for both guilds with a larger number of
iterations. The progressive ordering of contacts actually prevents this from
happening thus giving rise to a perfectly nested system with a different
order. A similar situation in which both guilds have different degree
distributions has been described \cite{Newman} for social webs such as films
and actors and boards and directors of large companies.

\subsection{Results for the projected graphs}

The interactions among mutualist species are blended into the weights and
strengths of the projected graphs. In what follows we discuss the results of
the corresponding distributions in comparison with empirically observed
data. These are shown in Figs.\ref{gradopr}, \ref{strengths} and \ref%
{dispesos}. All figures have the same organization of Fig.\ref%
{distribuciones}, namely upper panels correspond to animals (columns) while
lower panels correspond to plants (rows). Theoretical values are deduced
from the rectangular adjacency matrix whose degree distributions are shown
in Fig.\ref{distribuciones}. We show results that correspond to $R=1$, $%
R=0.1 $ and $R=10.0$. Empirically observed values are always displayed as a
reference in spite of the fact that values of $R$ that are different from 1
are not expected to represent biological networks.

\subsubsection{Degree and clustering distributions.}

\begin{figure*}[tbp]
\includegraphics[width=18cm]{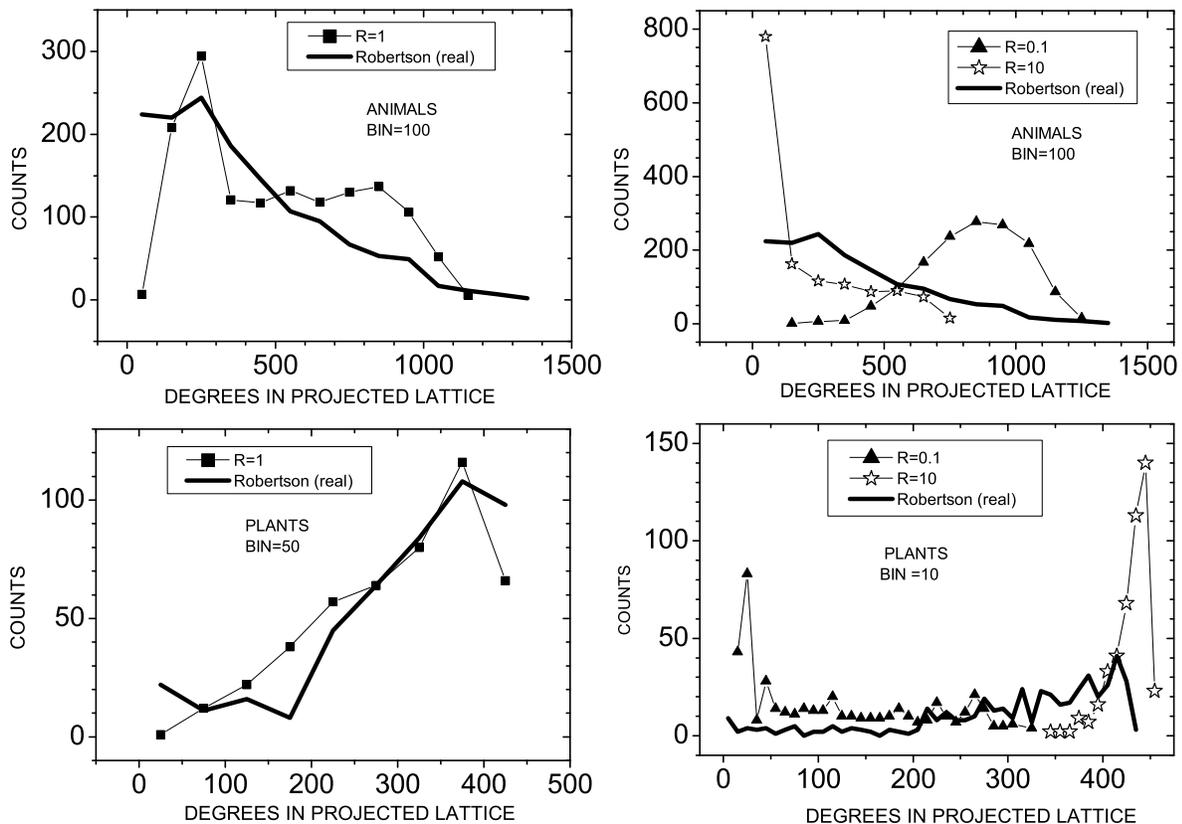}
\caption{{\protect\footnotesize Degree distributions in the projected graphs
for animals (upper panels) and plants (lower panels), for different values
of $R$. The shape of the full curves in the lower panels are diferent
because of the binning.}}
\label{gradopr}
\end{figure*}

In Fig. \ref{gradopr} we show the degree distributions of the projected
graphs for plants and animals. For a perfectly ordered system under Strategy
(I) and for any value of $R$ both distributions should approach a delta like
function located at the corresponding number of species. Since the
convergence to this limiting distribution is extremely slow, a partially
ordered system is expected to show significant departures from such extreme
distributions.

For $R\ll 1$ or $R\gg 1$ either columns or rows are updated more frequently
than rows or columns respectively. As explained above this causes one of the
two guilds to develop a prominent peak at the corresponding number of
species. This is the situation shown in the two right panels of Fig. \ref%
{gradopr} in which filled triangles and stars exchange roles showing a peak
at the extreme right.

For this same reason, the distributions for both guilds and $R=1$ are not
similar and indicate that one of the two guilds has reached a more ordered
configuration than the other. Indeed, while plants display a maximum at the
number of plant-species, animals have not yet developed such pattern showing
a maximum close to the origin. Since there are more columns than rows even
for $R=1$ each row is randomly selected for updating more frequently than
each column. As a consequence plants are closer to a situation of perfect
order. This effect is hard to observe directly in the bipartite graph.
Theoretical values obtained for $R=1$ are seen to closely reproduce the
empirically observed distributions while the SNM run with $R\ll 1$ or $R\gg
1 $ gives rise to projected graphs of a completely different nature.

We have also analyzed the distribution of clustering in the projected
graphs. The results obtained for this confirm those already shown for the
degree distributions. Upon a perfect nested order the distributions also
tend to be delta functions located at a maximum value of 1. This agrees with
the gradual approach to a tiny world pattern in which the projected graphs
are complete graphs, no matter the value of $R$. Such approach however
strongly depends upon the relative updating frequency of rows and columns.
When $R$ is widely different from 1, one of the two guilds shows a
distribution of clustering that is a bell shaped while the other guild
develops a strong peak at the maximum possible clustering equal to 1. For $%
R=1$ the distributions that are obtained are in good qualitative agreement
with the observed data in which both guilds appear not to be equally
ordered: while animals have a heavily skewed distribution with a maximum at
clustering equal to 1, plants have a more even distribution between
clustering 0.8 and 1.0.

\subsubsection{Distribution of strengths.}

\begin{figure*}[tbp]
\includegraphics[width=18cm]{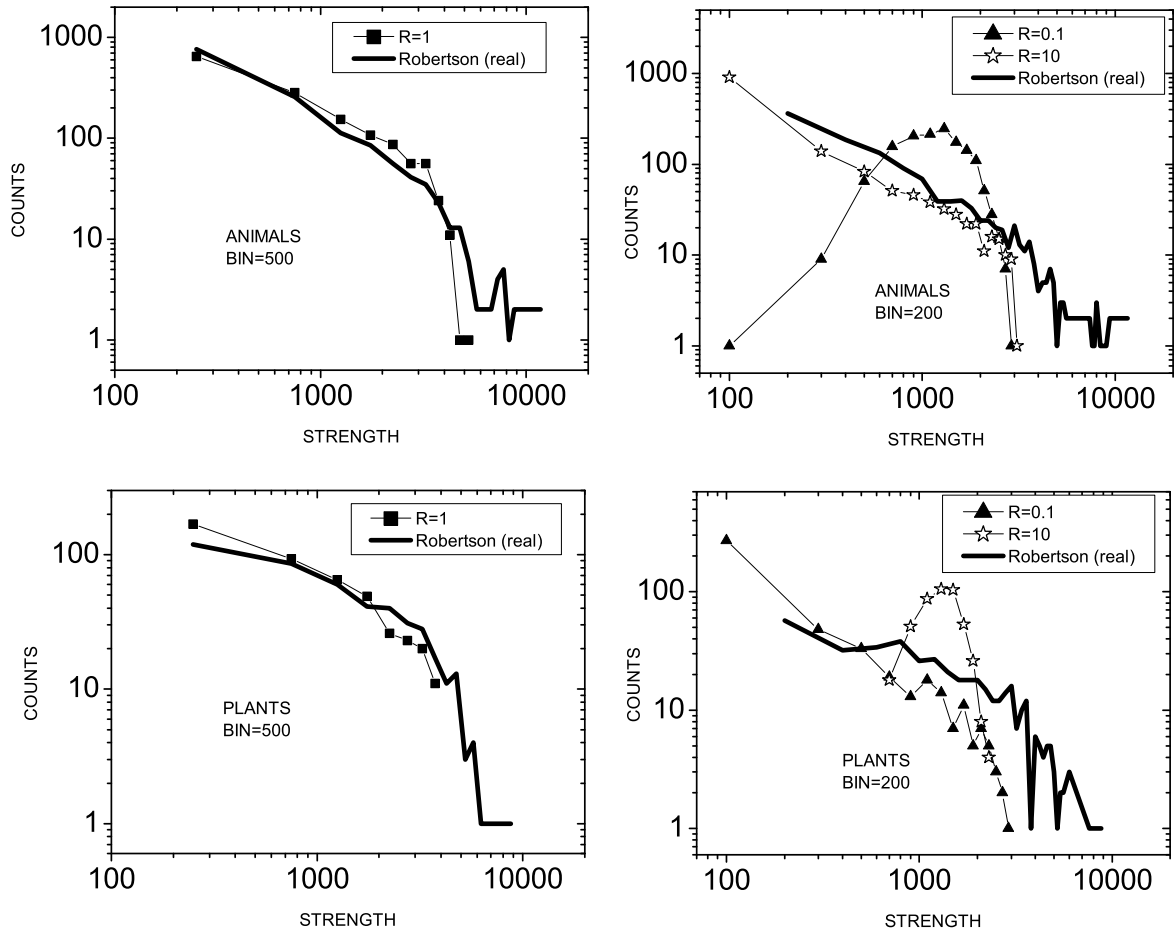}
\caption{{\protect\footnotesize Strength distributions in the projected
graphs for animals (upper panels) and plants (lower panels), for different
values of $R$..}}
\label{strengths}
\end{figure*}

The relative importance of the different species in both projected systems
is displayed by the distributions of strengths. These are shown in Fig.\ref%
{strengths}. These distributions provide insight about the relative
relevance of the different species in the two separate systems because they
combine into a single distribution the number of neighbors with the weights
of the interactions linking them. For a nested system with $R \simeq 1$,
distributions approach a (truncated) power law with a vast majority of nodes
having a low strength, close to their degree in the projected graph. On the
other hand, few nodes with many neighbors have a very high strength. If $R
\gg 1$ or $R\ll 1$ one of the two strength distributions is shaped as a bell
indicating little change with respect to the original random distribution,
while the other follows the pattern explained above. General features of
empirical values are also well described by the results obtained 
with the SNM (left panel of Fig. \ref{strengths})

\subsubsection{Distribution of weights.}

\begin{figure*}[tbp]
\includegraphics[width=18cm]{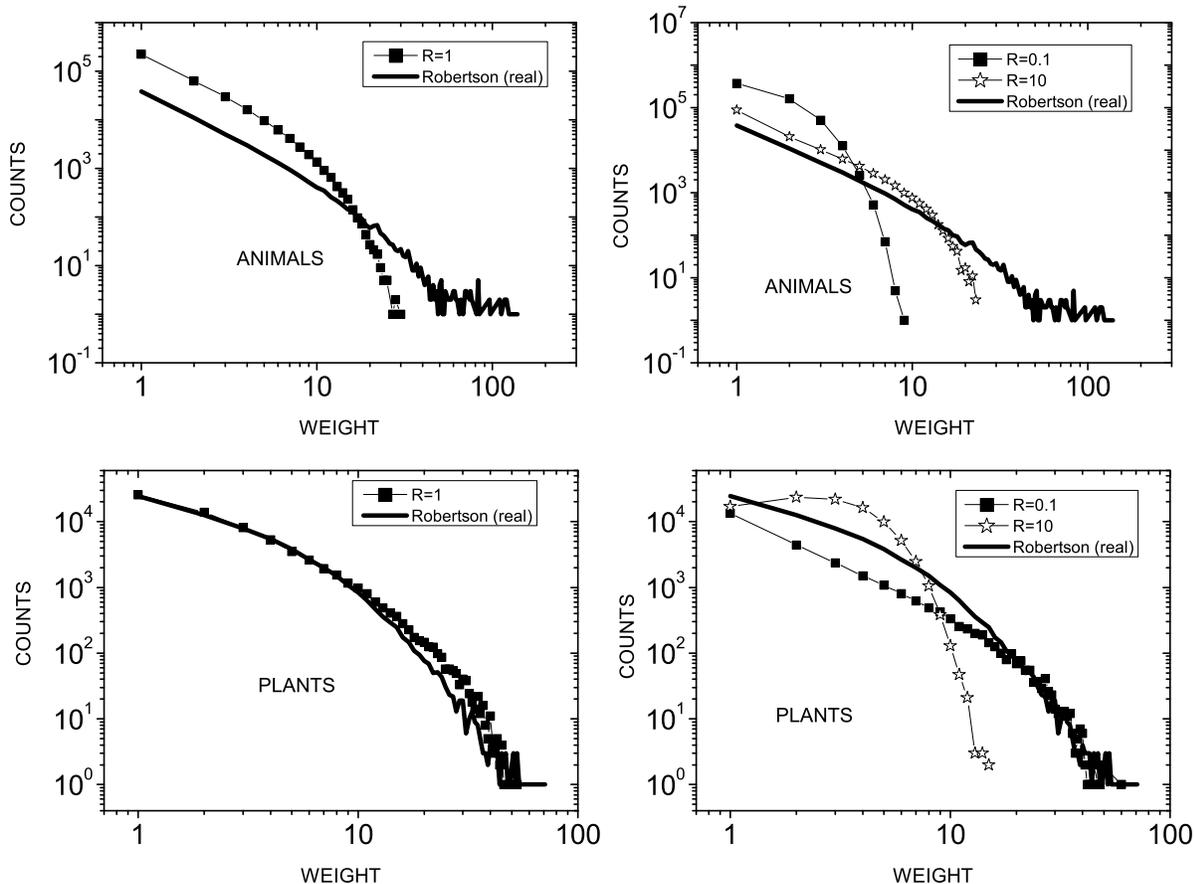}
\caption{{\protect\footnotesize Weight distributions of the links of the
projected graphs for animals (upper panels) and plants (lower panels), and
different values of $R$.}}
\label{dispesos}
\end{figure*}

As pointed up above, updating the interactions of one guild changes the
world that is seen by the other. The weights of the projected graphs
indicate how each guild mediates the relationship between individuals of the
other. Within the SNM such mediation changes with $R$. In Fig.\ref{dispesos}
we show the theoretical and the observed distribution of weights for plants
and animals with the same conventions of the preceding figures. In the limit
of perfect order and $R=1$, weights within the animal or plant systems
approach the decay of a truncated power law. The real 
system is not perfectly ordered
and therefore the distributions have a different decay rate. This is seen in
the two left panels of Fig.\ref{dispesos}. Real data for animals is seen to
have a distribution of weights that is closer to a power law than the
corresponding one of plants. This result is consistent with what has been
previously observed in connection with the degree distributions of the
projected graphs. The cases shown in the right panels confirm that values of 
$R$ that differ from 1 lead to a power-law distribution of strengths for
only one of the two guilds.

Although we have not attempted a detailed fit of empirical data, this is
seen to be qualitatively consistent with values $R\simeq 1$.

\subsubsection{Distribution of paths of minimum length.}

Besides the degrees, the strengths, the clustering coefficients, and the
weights of the links, the projected graphs are also described by the
distribution of the (minimal) path-lengths between any pair of nodes. This
distribution is strongly dependent upon the probability of contacts $%
\phi_{\pi}$ defined in Eq.(\ref{densipro}), between two species of the same
guild, that in turn depends upon the ordering process stemming from the SNM.

The distribution of minimal path lengths has a different pattern depending
upon the value of $R$ and upon the CPR that is used in the ordering process.
In Table (I) we show as an example the results obtained with the SNM in
several circumstances together with data of the observed system described in
Ref \cite{Robertson}. For social type networks i.e. when rows and columns of
the adjacency matrix are updated with very different frequencies, the
distribution of path lengths has a different shape for each guild. The less
frequently updated guild (in the example shown in Table (I) this corresponds
to the columns) has a larger fraction of longer paths.

The occurrence of longer paths can only take place through minute
alterations of a perfectly nested order of the bipartite network thus
turning the distribution of minimal path lengths into a powerful tool to
detect such alterations. This follows from the data of the network reported
in Ref.\cite{Robertson}. In this case one certainly checks the gross feature
of a great majority of shorter paths. Besides this, a small but significant
fraction of paths of length 3 is also present. This can not be expected to
result from fluctuations in the random initial conditions used in running
the SNM and is never reproduced for biologically sound values of the
parameters of the SNM. A way to understand this is the following. The 1.4 \%
of paths of length 3 are a total of almost 14,000 paths. This fraction of
longer paths can be obtained from the presence of a very small set of
perhaps a tenth of nodes that are linked by single vertices to a densely
interconnected core built up by the other nearly 1400 nodes that have
minimal paths only of lengths 1 and 2. Since the SNM is a statistical model
that deals equally with all species of both guilds one should not expect to
account for these sort of details.

\section{Conclusions}

We have presented a generalization of the SNM (Self-organizing Network
Model) that aims at finding the topology of a bipartite network that
optimally takes into account some local contact preference rule (CPR)
between the agents of the two guilds. To achieve this the rows and the
columns of the adjacency matrix alternatively and iteratively update their
contacts following such rule.

The CPR that favors contacts with a counterpart with as many links as
possible provided the best description of real mutualistic networks (Ref.%
\cite{nosotros1}). Thus an insect prefers to visit a flower that has many
other species visiting it and a flower tends to attract a wider variety of
insects. The opposite CPR that consists in developing some kind of
specialization by which contacts are as few as possible is not observed in
nature and gives rise to degree distributions that are sharply peaked at one
value.

We have also considered an extension of the SNM that amounts to change the
updating frequency of the contacts of both guilds. When this extra degree of
freedom is used, the model can also qualitatively account for the difference
between the degree distributions of the two guilds that is present in social
networks such as boards and directors. These results indicate that nested
bipartite networks may belong to two well differentiated classes. While $%
R\simeq 1$ represent the situation of ecological systems, $R\ll 1$ or $R\gg
1 $ correspond instead to social networks\cite{Newman}.

The distributions that have been observed in social networks indicates that
the CPR's are not the same for both guilds as in the case of mutualistic
webs. While the degree distribution of one guild (say the distributions of
the number of seats of a board) may follow some random process around a mean
that represents some common practice, the degree distributions of its
counterparts (directors) may instead be governed by a non random process
obeying some specific CPR similar to that of either plants or animals in a
mutualistic web. Thus, executives sitting in these boards may be the outcome
of a selection process based upon the special merit of sitting already in
several other boards (see Fig.8 of Ref.\cite{Newman}).

\begin{widetext}
\begin{center}
\textbf{Table 1} \\
\vspace{0.3cm}
\begin{tabular}{|l|c|c|c|c|c|l|}
\hline\hline
GUILD$^{(1)}$& $\phi_{\pi}(\%)$ & $D_1$ $^{(2)}$(\%)& $D_2$(\%)  & $D_3$(\%)  & $D_4$(\%) & COMMENTS \\
\hline\hline
ANIMALS & 26.6       & 26.7 & 71.9  & 1.4  & .004  & {\footnotesize Observed Ref.\cite{Robertson}} \\
\hline
PLANTS &  68.0       & 68.5 &  31.2 &  .3  &  -    &  {\footnotesize Observed Ref.\cite{Robertson}} \\
\hline\hline
ANIMALS & 22.1       &  22  & 78    &   -  &  -    & {\footnotesize Random} \\
\hline
PLANTS &  54.0       &  57  & 43    &   -  &  -    & {\footnotesize Random} \\
\hline\hline
ANIMALS & 32.9       &  32  & 68    &  -   &  -    & {\footnotesize 70,000 iter. of SNM $^{(3)}$}  \\
\hline
PLANTS  & 65.0       &  65  & 35    &   -  &  -    & {\footnotesize  70,000 iter. of SNM $^{(3)}$}\\
\hline\hline
ANIMALS & 77.1       &  77  & 23    &   -  &   -   & {\footnotesize 1,000,000 iter. of SNM $^{(3),(4)}$}\\
\hline
PLANTS  & 86.8       &  87  & 13    &   -  &  -    & {\footnotesize  1,000,000 iter. of SNM $^{(3),(4)}$}\\
\hline\hline
COLUMNS & 15.1       &  15  & 84    & .5   &  -    & {\footnotesize 70,000 iter. of SNM $^{(5)}$}\\
\hline
ROWS    & 89.5       &  90  & 10    &  -   &  -    & {\footnotesize  70,000 iter. of SNM $^{(5)}$} \\
\hline\hline
\end{tabular}
\end{center}
\end{widetext}

{\footnotesize Empirical distribution of minimal path lengths (first two
rows) for the system described in Ref.\cite{Robertson} and several results
of the SNM using Strategy I with an adjacency matrix of the same dimensions
and the same number of contacts.\newline
(1) Plants correspond to rows and animals to columns. \newline
(2) $D_j$ are the fraction of paths of length $j$ expressed as percentages
of the total number of paths. Theoretical results are rounded to the nearest
integer. \newline
(3) Plants and animals equally updated.\newline
(4) Bipartite network symmetrically nested, closer to perfect order than any
biological system.\newline
(5) Columns updated 10 times less frequently than rows. This simulation
would correspond to a social type bipartite network}

The presence of fat tails in biological mutualistic networks has
traditionally been attributed to the nested pattern of contacts between both
guilds. Indeed, when $R=1$ we have found that such is precisely the case.
However the present model indicates that there is not a close relationship
between the two concepts. With the present model \textit{any} bipartite
network that is adapted under the SNM \textit{always} reaches a perfectly
nested pattern of interactions independently of the updating frequency of
rows and columns. However, the degree distributions strongly depend upon the
different updating frequencies or, equivalently, by the prevailing order
within each guild. Thus nestedness has to be considered under a new light.
The particular degree distributions observed in mutualistic webs should be
attributed primarily to the particular way in which the pattern of
interactions is achieved or, to put it into different words, to the way in
which the CPR is actually enforced within the two guilds of the bipartite
web.

Another extension that we have considered is the study of the two projected
graphs and the changes that are induced on them by the SNM. We stress the
contribution of this study to the understanding of the differences between
social-type and ecological-type networks. In addition we have also checked
that the results of the SNM for the projected graphs with $R=1$ are also in
good qualitative agreement with empirical observations of real biological systems.
Moreover the projected graphs can reveal if plants and animals have 
achieved a different degree of order, something that is difficult to observe 
directly from the bipartite matrix.

We have studied the distribution of minimal path lengths within the
projected graphs. This distribution is strongly dependent upon the
probability of contacts. Since this is a rapidly growing function of the
density of contacts in the bipartite matrix, the projected systems tend to
always be tiny worlds of very densely interconnected species. Moreover, the
theoretical estimates of the distribution of minimal path lengths for the
system of ref.\cite{Robertson} shed some light on possible reasons of minute
departures from perfect order. The lack of a small fraction of paths of
length 3 may perhaps be taken as a hint of the presence of species within
the system having different CPR's that, in turn may produce within the
ecosystem what in ecology are called \textit{compartments} (\textit{%
communities} in complex network language), i.e. groups of species that while
strongly connected among themselves, are weakly connected to the rest of the
network.


\begin{thebibliography}{99}
\bibitem{pie2} The states that are reached in this way may however not have
all contacts placed below the isocline of perfect order. One can construct a
pattern of interactions that acts as a trap -i.e. a local minimum -
stagnating the ordering process. Thus if during the iterative process the
system falls into one of these configurations it will not undergo any
further reorganization and yet will not correspond to an optimal allocation
of contacts. Such traps could be avoided for instance by mapping the
self-organization process into the solution of an optimization problem
through simulated annealing. The occurrence of these traps is however not
relevant to our present analysis because they are not important when dealing
with real biological webs that are usually far from an asymptotically
ordered pattern. If one aims at fitting empirical data the self-organizing
process has therefore to be stopped at a moderate number of iterations. We
will therefore ignore this problem for the moment.
\bibitem{pie3} In this particular case, the asymptotically ordered state may
not be a static configuration. Some nodes may exchange links keeping the
same average degree. This however does not change any statistical
consideration.
\bibitem{piex} In the rest of the work we only discuss degree distributions and not \textsl{%
cumulative} degree distributions such as it is a usual practice in the
biological or ecological communities.
\end{thebibliography}
\end{document}